# Harnessing AI data-driven global weather models for climate attribution: An analysis of the 2017 Oroville Dam extreme atmospheric river

Jorge Baño-Medina[a], Agniv Sengupta[a], Allison Michaelis[b], Luca Delle Monache[a], Julie Kalansky[a], Duncan Watson-Parris[a,c]

[a] *Center for Western Weather and Water Extremes, Scripps Institution of Oceanography, University of California San Diego*

[b] *Department of Earth, Atmosphere, and Environment, Northern Illinois University, DeKalb, IL*

[c] *Halıcıoğlu Data Science Institute, University of California San Diego, La Jolla, CA, USA*

*Corresponding author*: Jorge Baño-Medina, jbanomedina@ucsd.edu

## ABSTRACT

AI data-driven models (Graphcast, Pangu Weather, Fourcastnet, and SFNO) are explored for storyline-based climate attribution due to their short inference times, which can accelerate the number of events studied, and provide real time attributions when public attention is heightened. The analysis is framed on the extreme atmospheric river episode of February 2017 that contributed to the Oroville dam spillway incident in Northern California. "Past" and "future" simulations are generated by perturbing the initial conditions with the pre-industrial and the late-21st century temperature climate change signals, respectively. The simulations are compared to results from a dynamical model which represents plausible "pseudo-realities" under both climate environments. Overall, the AI models show promising results, projecting a 5-6 % increase in the integrated water vapor over the Oroville dam in the present day compared to the pre-industrial, in agreement with the dynamical model. Different geopotential-moisture-temperature dependencies are unveiled for each of the AI-models tested, providing valuable information for understanding the physicality of the attribution response. However, the AI models tend to simulate weaker attribution values than the "pseudo-reality" imagined by the dynamical model, suggesting some reduced extrapolation skill, especially for the late-21st century regime. Large ensembles generated with an AI model (>500 members) produced statistically significant present-day to pre-industrial attribution results, unlike the >20-member ensemble from the dynamical model. This analysis highlights the potential of AI models to conduct attribution analysis, while emphasizing future lines of work on explainable artificial intelligence to gain confidence in these tools, which can enable reliable attribution studies in real-time.

## SIGNIFICANCE STATEMENT

Large volumes of greenhouse gasses are continuously emitted into the air, warming the Earth's atmosphere, and increasing the frequency and intensity of extreme events, such as heat waves, droughts, and heavy precipitation. Climate attribution by means of the storyline approach aims to quantify the human fingerprint in observed extreme events to establish a connection between climate change and the weather we experience. Attribution studies currently rely on dynamical models to simulate the climate system, which require huge computational resources and demand long simulation times, limiting the number of events analyzed and preventing reports to be published in near-real time, where public attention is at its peak. Here, AI data-driven models are examined for this task.

## 1. Introduction

Large volumes of greenhouse gasses (GHG) are continuously emitted into the air, warming the Earth's atmosphere (IPCC, 2022), and increasing the frequency and intensity of extreme events, such as heat waves (Russo et al., 2014, Guo et al., 2017, Dosio, 2017, Molina et al., 2020), droughts (Cook et al., 2020, Vicente-Serrano et al., 2022), and heavy-rains (Trenberth, 2005; Frei et al., 2006, John et al., 2022). The field of climate attribution aims to quantify the human fingerprint on climate, and looks for establishing connections between the weather we experience and climate change.

Different climate attribution methodologies exist in the literature (see e.g., Hulme, 2014, and Shepherd 2016, for a review). In this study we focus on the storyline approach (Hoerling et al., 2013). This approach compares sets of simulations, which are framed in different emission scenarios of GHG (e.g., the "present" and the "past"). For the former, the model is initialized using reanalysis data, whereas the latter translates the initial condition field to the "past" climate by removing the climate change signal, or "delta", of the temperature variables. Several studies have adopted this methodology to quantify the human fingerprint on the intensity of an observed extreme event (e.g., Rasmussen et al., 2011; Mallard et al., 2013; Hoerling et al., 2013; Hazeleger et al., 2015; Lackmann, 2015; Trapp & Hoogewind, 2016; Shepherd, 2016). Similarly, analysis for plausible future climates can be undertaken to evaluate how certain extreme events may intensify in the future (e.g., Michaelis et al., 2022). Storyline approach simulations are currently run with either Global (GCMs) or Regional Climate Models (RCMs; Hegerl & Zwiers, 2011), which are the main tools used to study the evolution of climate. However, their huge computational costs and long simulation times limit the number of events that are currently analyzed, and prevent these experiments from being applied in near-real time, where public attention is peaked. Furthermore, forecast uncertainty, which can arise from imperfections in both the model and/or in the initial conditions, is usually under-sampled due to the small number of members in the ensemble forecasts. This undersampling is extremely relevant since climate attribution studies usually deal with extreme events where the characterization of uncertainty is of utmost importance.

In this study, AI data-driven models that have generated substantial research interest in the past couple of years, especially in the domain of weather forecasting, are examined for climate attribution following the storyline approach. These are novel tools based on deep learning (see Goodfellow et al., 2016), in particular graphical neural networks and vision transformers, that have achieved impressive deterministic (Pathak et al., 2022, Bi et al., 2023, Lam et al., 2023), and probabilistic (e.g., Baño-Medina et al., 2024a) forecast skill on par with state-of-the-art Numerical Weather Prediction (NWP) models. They are trained with large volumes of data, typically global reanalysis datasets, to learn a function between consecutive atmospheric time-steps. Thus, given a set of input features, e.g., global atmospheric fields at 0.25° of spatial resolution, they forecast the next weather state (defined by the same set of variables) 6-hours ahead. To produce forecasts at longer lead times, the prediction is used as input to the next iteration (auto-regressively). Once trained, the generation of multi-member forecasts from such AI models is orders of magnitude faster and requires significantly less energy compared to NWP models. Here we explore the potential of these AI data-driven models to support climate attribution studies for extreme events, and demonstrate their suitability for near-real time applications given their ability to generate a very large number of ensemble members in a relatively short time to better capture the uncertainty.

The case study selected is the Oroville dam atmospheric river (AR) episode of February 2017 (hereafter just Oroville AR). ARs are long, narrow corridors of intense water vapor transport in the lower troposphere (Zhu and Newell, 1998; Ralph et al., 2004, 2005, 2017), and are linked to several hydrometeorological impacts over the mid-latitude regions, including beneficial outcomes like alleviation of droughts or mitigation of wildfire risks (Ralph et al., 2019), as well as severe hazards like debris flows (Oakley et al., 2018; McGuire et al., 2021), flooding (Corringham et al., 2019) and extreme wind (Waliser and Guan, 2017). Climate change simulations project more intense and more frequent ARs, as a consequence of a warmer atmosphere that accepts larger volumes of moisture, leading to more frequent and more intense precipitation (Gao et al., 2015; Espinoza et al., 2018; Gershunov et al., 2019; Payne et al., 2020; Rhoades et al., 2020; Zhang et al., 2021; Baek & Lora, 2021). In particular, the Oroville AR was an impactful extreme event that contributed to damage at Oroville dam (California, United States) leading to the evacuation of over 150,000 people (Vano et al., 2018; White et al., 2019; Henn et al., 2020). The Oroville AR has already been examined by means of the storyline approach using a dynamical model (Michaelis et al., 2022), attributing 11–15% of the AR's precipitation to human action, and projecting future enhancements upwards of 21–59% under future climate conditions. Using dynamical simulations, such as those of Michaelis et al. (2022), as a "groundtruth/pseudo-reality" is a common approach in the evaluation of statistical models, especially for climate change analysis (Vrac et al., 2007; Baño-Medina et al., 2021; Baño-Medina et al., 2022; Balmaceda et al., 2023) where future observations are unavailable. Doing so allows us to test key properties of the statistical model, such as, the ability to extrapolate to warmer/cooler climates compared to their training data, or the physical response to perturbations in the temperature fields (i.e., "climate-change" perturbations).

## 2. Data

The European Centre for Medium-Range Weather Forecasts (ECMWF) Reanalysis version 5 (ERA5, Hersbach et al., 2020) currently provides the most accurate representation of the recent historical atmosphere. This dataset assimilates observational measurements of weather with "first-guess" forecasts to produce global atmospheric fields for a large number of variables at 0.25° of spatial resolution. The AI-models of this study were trained on ERA5 using values at 6-hour intervals (0, 6, 12, 18 UTC) for the 1979–2015 period, and are publicly available at the ECMWF Github repository (see Data availability section). The Integrated Water Vapor (IWV) from ERA5 is downloaded from the ECMWF portal and used as "groundtruth" to evaluate the AI-models in the present climate. This variable, representing the total amount of water vapor (precipitable water) within a square meter column of air, is an important descriptor of ARs (Reid et al., 2020).

Temperature climate change signals, or "deltas", were derived from a subset of 20 GCMs from the Coupled Model Intercomparison Project Phase 5 (CMIP5; see Table 2 in Michaelis et al. (2019) for the complete list of GCMs). GCMs simulate the global climate based on different emission scenarios called Representative Concentration Pathways (RCP, Meinshausen et al., 2011), including historical, pre-industrial, and future periods. The future "deltas" were computed by taking the difference between the February climatological mean temperatures during the late 21st century (2080–2099) following the high-emissions RCP8.5 and the historical (1980–1999) period. Similarly, the "past" deltas were computed considering the pre-industrial mean temperatures (1880–1899) in comparison to the historical period. While the RCP8.5 simulation presents the strongest climate change signal, and therefore allows us to evaluate the extrapolation ability of the AI models, the pre-industrial scenario represents a case of moderate extrapolation, and exemplifies the

File generated with AMS Word template 2.0

baseline against which the human fingerprint is measured for observed current events. While CMIP6 model data is now available, CMIP5 is preferred for this study for direct comparison with Michaelis et al., 2022.

## 3. Methods

*a. Model Prediction Across Scales-Atmosphere (MPAS-A)*

The Model Prediction Across Scales-Atmosphere (MPAS-A; Skamarock et al., 2012) is a global, atmosphere-only, nonhydrostatic numerical model. MPAS-A integrates the atmospheric differential equations over a variable-resolution latitude-longitude grid that builds on unstructured Voroni meshes (Du et al., 1999), where spatial resolution gradually coarsens proportionally to the distance to a centroid. These types of models are well-versed in simulating regional climates and provide a more complete description of the global atmospheric state than limited-area RCMs (Park et al., 2014). In particular, Michaelis et al. (2022) ran their MPAS-A simulations using a 3–60 km variable resolution mesh with the 3-km centroid on 37°N, 126°W at 3-hourly temporal resolution. This mesh orientation produced 3-km fields over California, and extended about 15-km offshore into the Pacific Ocean, where the AR originated.

MPAS-A contains an initialization module that adjusts the geopotential, winds, pressure variables, and specific humidity to the thermodynamic state, in case this one is modified. This component of the model is key, since for the storyline approach the temperature variables are added to the initial condition (see Section 4).

*b. Adaptive Fourier Neural Operators (AFNO)*

The Adaptive Fourier Neural Operator (AFNO, Guibas et al., 2021; Pathak et al., 2022), is a neural network that performs self-attention in the space of frequencies by means of the Fast Fourier Transform (FFT) and a Vision Transformer backbone (ViT, Dosovitskiy et al., 2020). The latter combination was found capable of dealing with high-dimensional input spaces, where the large number of parameters in the network made self-attention simply intractable. This topology was tested for weather forecasting in Pathak et al. (2022), being the first AI model to ever achieve comparable results to NWP systems. Given a set of input features the model was tasked with learning the atmospheric state 6-hours ahead, using 8 AFNO blocks, thus repeatedly transforming and de-transforming from the Fourier space to the latitude and longitude grid, at deeper levels of complexity as the operations flow deeper in the network.

This model uses the following set of 20 input variables: surface air temperature; mean sea level pressure; surface pressure; integrated water vapor; surface zonal and meridional wind; relative humidity and air temperature at 500 and 850 hPa; zonal and meridional wind at 500, 850 and 1000 hPa; and geopotential at 50, 500, 850 and 1000 hPa. The reader is referred to Pathak et al. (2022) for more details on AFNO.

*c. Ensemble of Adaptive Fourier Neural Operators (EnAFNO)*

The Ensemble of Adaptive Fourier Neural Operators (EnAFNO) is a strategy to produce ensemble forecasts by means of six bred vectors and 90 AFNO models, resulting in a 540-member ensemble (Baño-Medina et al., 2024). This ensemble achieved probabilistic skill on par with state-of-the-art NWP systems for the IWV, and showed an impressive skill to reproduce the main features of an extreme AR over California. While bred vectors are estimations of initial condition uncertainty (Toth and Kalnay, 1997), the 90 AFNOs represent the uncertainty in the model parameters. Models are retrieved during the training phase by sampling the coefficients at the different epochs with a modified learning rate schedule. Bred vectors are computed by generating 6 different breeding cycles—starting January 26th at 12 UTC—by adding gaussian noise to ERA5 (perturbed forecast). A control forecast, fed with non-perturbed ERA5 data, is removed from the perturbed one, and the differences, i.e., the bred vector at time *t+6* hours, are scaled to match the norm of the gaussian noise. The bred vector is then added to the ERA5 data at time *t+6*, and the process is repeated. Each breeding cycle requires about 8-12 iterations −1-2 days of lead time– until it converges to estimations of initial condition uncertainty. Then, these pre-computed bred vectors are used to perturb the initial condition of the forecast. The reader is referred to Baño-Medina et al., 2024 for more details on the ensemble generation strategy, and the breeding cycle. EnAFNO uses the same set of 20 input variables as AFNO of Pathak et al., 2022.

*d. Spherical Fourier Neural Operator (SFNO)*

The Spherical Fourier Neural Operator (SFNO; Bonev et al., 2023) is an updated version of AFNO, also tasked to forecast the next 6 hours of weather. This model uses spherical harmonics instead of the FFT, therefore providing a better representation of the Earth's sphere. It uses the following set of 73 input variables: surface air temperature; mean sea level pressure; surface pressure; integrated water vapor; surface zonal and meridional wind; zonal and meridional wind at 100 meters; relative humidity, air temperature, zonal and meridional wind, and geopotential, at the following pressure levels – 50, 100, 150, 200, 250, 300, 400, 500, 600, 700, 850, 925 and 1000 hPa.

*e. Pangu-Weather*

Pangu-Weather (hereafter just Pangu), builds on 3D Earth-specific transformer (3DEST) blocks to model the Earth's geometry (Bi et al., 2023). First, a patch embedding layer maps the input fields to a smaller dimension. Second, the resulting (reduced) 3D (for pressure variables) and 2D (for surface variables) fields feed an encoder-decoder Swin transformer (Liu et al., 2021), with 8 layers each, where each layer is a 3DEST block. The positional bias of the transformer is set to consider the fact that distances between adjacent grid-points vary with latitude. Finally, a patch recovery layer maps the decoder outputs to the global 3D and 2D dimensions representing the weather state forecast 6-hours ahead.n Bi et al. (2023), different versions of Pangu were trained to forecast the next weather states at different lead times (i.e., 1-hour, 3-hours, 6-hours, 24-hours), however for consistency with the other AI models, only Pangu-6 is utilized here. Pangu uses the same set of variables as SFNO only without the integrated water vapor.

*f. Graphcast*

Graphcast is a Graph Neural Network (GNN) tasked to forecast the next 6 hours of weather (Lam et al., 2023). GNNs are generalizations of the rigid Convolutional Neural Networks (CNN; LeCun et al., 1995) to heterogeneous gridded data (e.g., Scarselli et al., 2008), where a set of learnable weights are only allowed to convolute over a 2-D equally-spaced gridded data. Differently, GNNs build on graphical structures, which are defined by a set of nodes and edges. Each node is a representation of the features, e.g., atmospheric variables at a certain gridpoint, and the edges establish connections between them, making possible the design of complex topological structures. GNNs are trained to learn a set of weights that transforms the input graph into another representation with the same graphical structure by leveraging the features at each node with information from its neighboring ones (and itself). This is called *message-passing*. By performing message passing multiple times, the information of one node can flow through the graph exchanging information to those located further. In particular, Graphcast consists of three modules: an encoder, a processor, and a decoder. First, an encoder maps the latitude-longitude grid of input data to a multimesh high-resolution graph defined by refining an icosahedron multiple times. This multi-mesh graph represented by the different icosahedrons allow modeling interactions between nodes at different spatial scales, from local connections to long-range ones (see Lam et al., 2023). Second, a processor performs repeated steps of message-passing through the different meshes of the icosahedrons. Third, a decoder maps back each node of the multi-mesh graph to the original latitude-longitude grid.

Graphcast uses the same set of variables as Pangu, but with one key difference. As opposed to the other AI data-driven models appearing in this study, which use a set of variables at time *t* to forecast time *t+6 (hours)*, Graphcast uses also information from the previous time, such that the forecast at time *t+6*, depends on the values at time *t* and also time *t-6*.

*g. NN-INIT*

The NN-INIT network is an AFNO topology trained with 6-hourly data for the period 1979–2015 from ERA5, to emulate the MPAS-A initialization module. This network uses as input the temperature and relative humidity fields at every available pressure level at time *t* and is tasked to minimize the mean squared error of the IWV, geopotential, surface pressure and mean sea level pressure at the same time *t*. The set of input and output variables was decided based on the MPAS-A initialization module configuration, where relative humidity was held constant at initial time. The model was trained on a cluster of 64 P100 GPUs using an Adam optimizer with a learning rate of 5E-4 and a cosine annealing scheduler for 100 epochs.

**4. Model simulations**

Following Michaelis et al. (2022), the goal is to simulate the Oroville AR (from February 5th, 2017 at 12 UTC to February 11th, 2017 at 12 UTC) based on different climate scenarios: the "present", the "past", and the "future". For the "present" simulation, the models are initialized on February 5th, 2017 at 12 UTC using data from ERA5. For the "past" and "future" scenarios the initial condition is modified following two steps (see Figure 1 for the AI-models). First, the temperature climate change signals for the month of February at all pressure-, sea-, and deep soil-levels available in each model are added to the initial conditions. Second, the modified initial condition feeds the initialization module, whose goal is to adjust the atmosphere to the new thermodynamic state. For this, the network NN-INIT estimates the IWV, mean sea level pressure, surface pressure, and geopotential at 50, 500, 850, and 1000 hPa. In parallel, the geopotential fields are adjusted to be in hydrostatic balance with the modified temperature fields at initial time (Jiménez-Esteve et al., 2024), using the temperature and relative humidity fields at each pressure level (see Appendix A). This is usually a good assumption of the Earth's atmosphere for hourly time scales and longer (Merriam, 1992). Finally, the remainder of the variables are concatenated, and the resulting initial condition is fed to the model.

The impact of the initialization module is analyzed in Section 5.d , and simulations for the AI-models without this component are also run. This allows us to explore whether or not they are able to physically adjust the other variables under a warmer/cooler atmosphere, i.e., respect the temperature-moisture link. This aspect is directly linked to their intrinsic ability to learn inter-variable dependencies in the atmosphere, which has not been deeply explored, and is key to climate attribution studies.

For the probabilistic forecasts, MPAS-A breeds from the 21 initial conditions from the Global Ensemble Forecast System (GEFS), while 6 breeding cycles per AFNO within the EnAFNO system are generated, starting January 26th, 2017 at UTC from a gaussian noise with a scale factor of 0.15. By starting the breeding cycle a few time steps earlier than the initial condition, the bred vectors are allowed to shift from the original gaussian noise (spin-up).

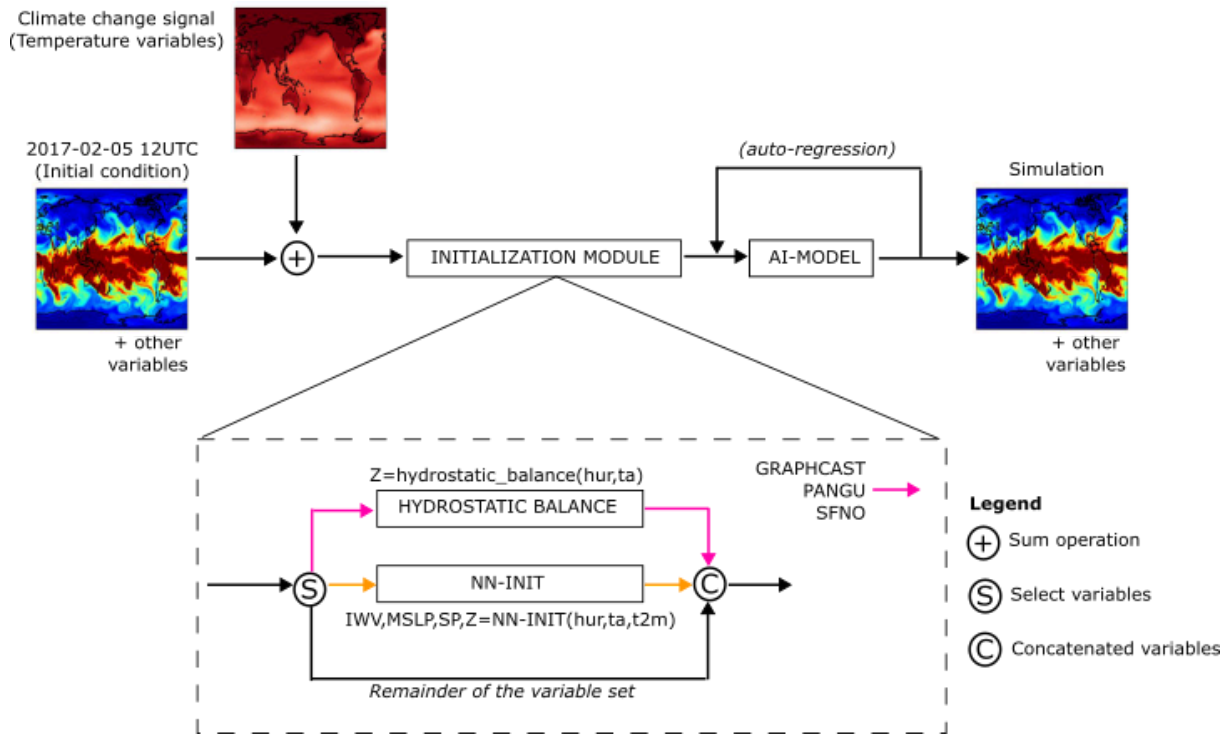

Fig. 1. Schematic representation of the framework to conduct attribution analysis with AI-models. The relative humidity (r), and the air temperature (ta) at every pressure level are used to adjust the geopotential (z) for Graphcast, Pangu and SFNO (Appendix A). Differently, for the AFNO the geopotential is computed using the NN-INIT network, since this model presents very few pressure levels (<4). The Integrated Water Vapor (IWV), the mean sea level pressure (MSLP), and the surface pressure (SP) at initial time are also estimated by means of the NN-INIT network using as input fields the relative humidity, the air temperature, and the 2-meter temperature (t2m). Note that for Graphcast and Pangu the IWV field is computed using the integration scheme described in Appendix B.

## 5. Results

*a. Forecasting the Oroville AR*

Simulations of the four aforementioned AI data-driven models (third to sixth rows) and MPAS-A (second row) are compared to ERA5 ("groundtruth"; first row), and displayed in Figure 2. Columns show the evolution of the IWV over the North Central and North Eastern Pacific Ocean and Western coast of the United States (US) for four different dates: February 5th at 12 UTC, February 6th at 12 UTC (lead time of 24 hours), February 7th at 12 UTC (lead time of 48 hours), and February 9th at 00 UTC (lead time of 84 hours). These dates represent important episodes of the event: the initial condition, the end of MPAS-A spin up in Michalis et al. (2022), the first peak, and the second peak of the AR. The Oroville AR moved eastward from the Central Pacific to the US transporting more than 40 kg/m² of moisture per
11

gridpoint from the tropics to mid latitudes. The first peak of the AR reached the Western US about 36-hours later with the highest values of IWV at 48-hours over the Oroville dam location (yellow box). Concurrently, a second AR merged with the first one, producing a longer second peak and more intense than the previous one, with maximums of IWV over the Oroville dam at 84-hours of lead time. These main features are also reproduced in the forecasts of the AI data-driven models and the MPAS-A, with some differences. Contrary to AFNO and SFNO, Graphcast and Pangu do not contain the IWV in their variable set. For these models, IWV is retrieved by integrating the specific humidity over 13 pressure levels (see Appendix B). The errors obtained in these models may be attributed to the limited number of available pressure levels, where the AR shows overall higher IWV values compared to ERA5 at every time frame analyzed. This is also evident in the fields at initial time, where both Pangu and Graphcast exhibit errors of 6.77 kg/m² over the Oroville dam. Differently, the MPAS-A shows an error of 1 kg/m² after interpolation to a 0.25° grid to facilitate a direct comparison with the AI models. For the two peaks observed in the Oroville AR, MPAS-A shows lower errors than the AI models over its duration, but most importantly all models, both dynamical as well as AI, are able to predict the main evolution and characteristics of the two peaks.

*b. On the extrapolation of AI data-driven models*

AI data-driven models are trained on historical records and they lack explicit physics in the formulation, therefore their extrapolation abilities are not yet clear. Towards addressing this gap, two different cases of extrapolation are analyzed: a strong and a moderate one, building on the "future" and "pre-industrial/past" scenarios, respectively. Figure 3 displays the differences between the "future" and "present" simulations, and between the "present" and "past" ones, for the air temperature at 850 hPa. For the strong extrapolation case, results are shown for the same dates considered in Figure 2; for simplicity, for the moderate case, we show only the second peak (last column). The "delta" fields for both cases are located in the top-left panelside. Here, the dynamical simulations with MPAS-A are used as a reference (or "pseudo-reality") to measure the extrapolation skill of the AI models, following a well-established evaluation practice in climate change studies (e.g., Vrac et al., 2007). Therefore, the closer the AI data-driven models are to the MPAS-A results the better the extrapolation skill. The temperature "delta" fields added to the initial condition of MPAS-A, evolve with time preserving the signal over most of the spatial domain, with only very small areas showing no signal or even negative values. The latter might be caused by differences in the location of various atmospheric features due to the applied temperature changes. Pangu and Graphcast lose part of the signal, as compared to MPAS-A, but overall the temperature change pattern is broadly preserved through the entire AR episode. Meanwhile, AFNO, and especially SFNO, loses almost the entire signal at 24-hours of lead time, showing little-to-no extrapolation skill. Similar conclusions are derived from the moderate extrapolation case, where the AI data-driven models show weaker signals than MPAS-A, at the second peak and the remainder of the AR episode (not shown), especially over land.

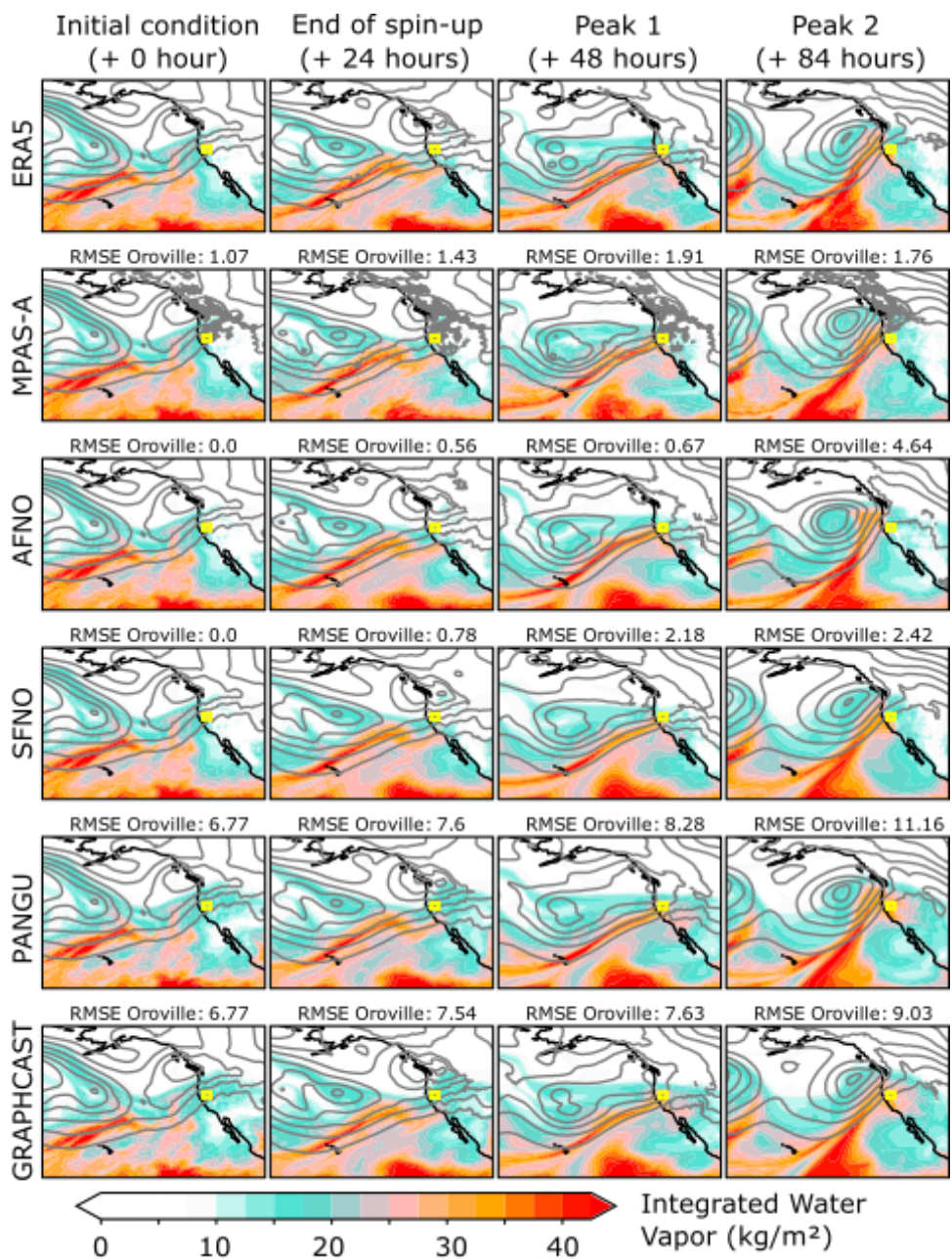

Fig. 2. ERA5, and "present" simulations of MPAS-A and the AI data-driven models for the Oroville's AR. Columns show the evolution of the Integrated Water Vapor (IWV) over the North Central and North Eastern

Pacific Ocean and Western coast of the United States (US) for four different dates: February 5th at 12 UTC, February 6th at 12 UTC (lead time of 24 hours), February 7th at 12 UTC (lead time of 48 hours), and February 9th at 0 UTC (lead time of 84 hours). These dates represent important aspects of the event: the initial condition, the end of MPAS-A spin up in Michaelis et al., 2022, the first peak, and the second peak of the AR. Contours display the geopotential at every 500 m²/s².

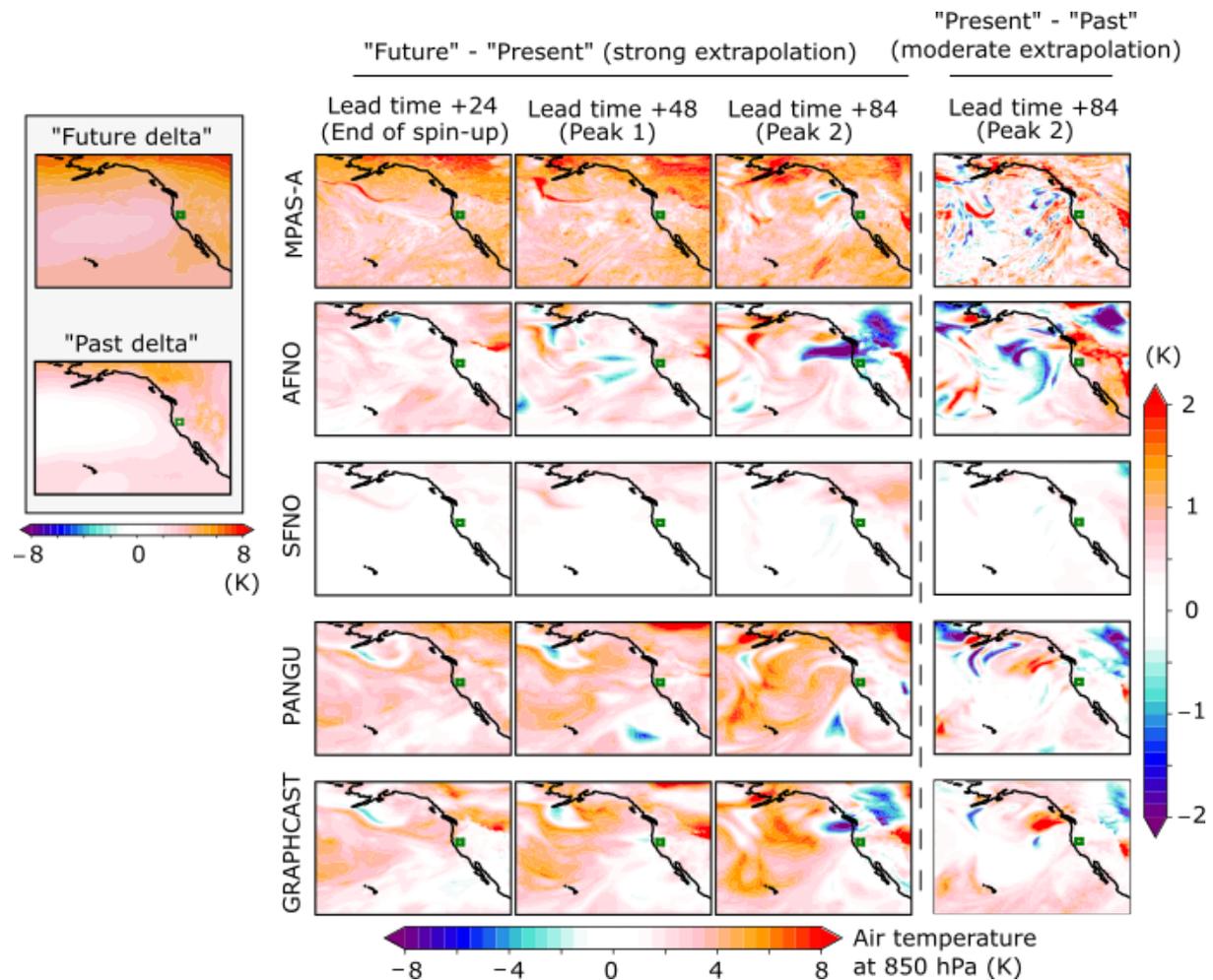

Fig. 3. "Future" minus "present" (strong extrapolation) simulations of MPAS-A and the AI data-driven model of air temperature at 850 hPa over the North Central and North Eastern Pacific Ocean and Western coast of the United States (US) for three different dates: February 6th at 12 UTC (lead time of 24 hours), February 7th at 12 UTC (lead time of 48 hours), and February 9th at 0 UTC (lead time of 84 hours), representing the end of MPAS-A spin up in Michaelis et al., 2022, the first peak, and the second peak of the AR, respectively. "Present" minus "past" (moderate extrapolation) simulations are displayed in the last column for the second peak. The far-future and pre-industrial "delta" fields of the air temperature at 850 hPa are shown in the top-left side. The colorbar on the left- and right-side describe the magnitudes for the strong, and moderate extrapolation cases, respectively.

*c. On the moisture response to "climate-change" perturbations*

AI data-driven models are not explicitly tasked to learn the interactions between atmospheric variables. Instead, these interactions are assumed to be indirectly learned during model training. Therefore, how perturbations in an atmospheric field modifies other variables in AI data-driven models is not yet clear. For this case of the Oroville AR, the moisture-temperature link is explored to investigate the realism of other atmospheric variables with perturbations in the initial condition temperature fields. Figure 4 displays the signals of IWV through the aforementioned four dates of interest for the MPAS-A and AI models. Again, for the moderate extrapolation case, only the second peak is shown. MPAS-A shows increased moisture, consistent with a warmer atmosphere. The highest increments are located in the tropics and in the AR at every time step, surpassing the values of 10 kg/m² at every gridpoint. Pangu and Graphcast also project increased values of moisture from the very beginning of the AR through the very end of the episode, with magnitudes comparable to MPAS-A for most of the domain. The spatial patterns of the moisture differences are also similar between MPAS-A and select AI models. For example, the arrow-shape negative values (in blue) located southeast of Alaska, in the MPAS-A during the second peak are also observed for Pangu, Graphcast, and AFNO, albeit with different intensities. These properties might be indicative of the physically consistent response in the IWV field to temperature changes in the initial condition. Finally, SFNO projects virtually unaltered values of moisture, consistent with the loss of temperature signal (Figure 3). AFNO also shows increases in moisture, with remarkable similarities with the spatial pattern of the Graphcast simulation, e.g., the negative values observed just north of the Oroville dam.

For the moderate extrapolation case, the AI-models show consistent results with MPAS-A, showing increases of the filament of the AR over the ocean and also in the tropics (most southern part in the panels). However, some lack of extrapolation is still diagnosed for this case, especially over land for Graphcast and Pangu. A detailed analysis over the Oroville dam domain is shown in Figure 6.

File generated with AMS Word template 2.0

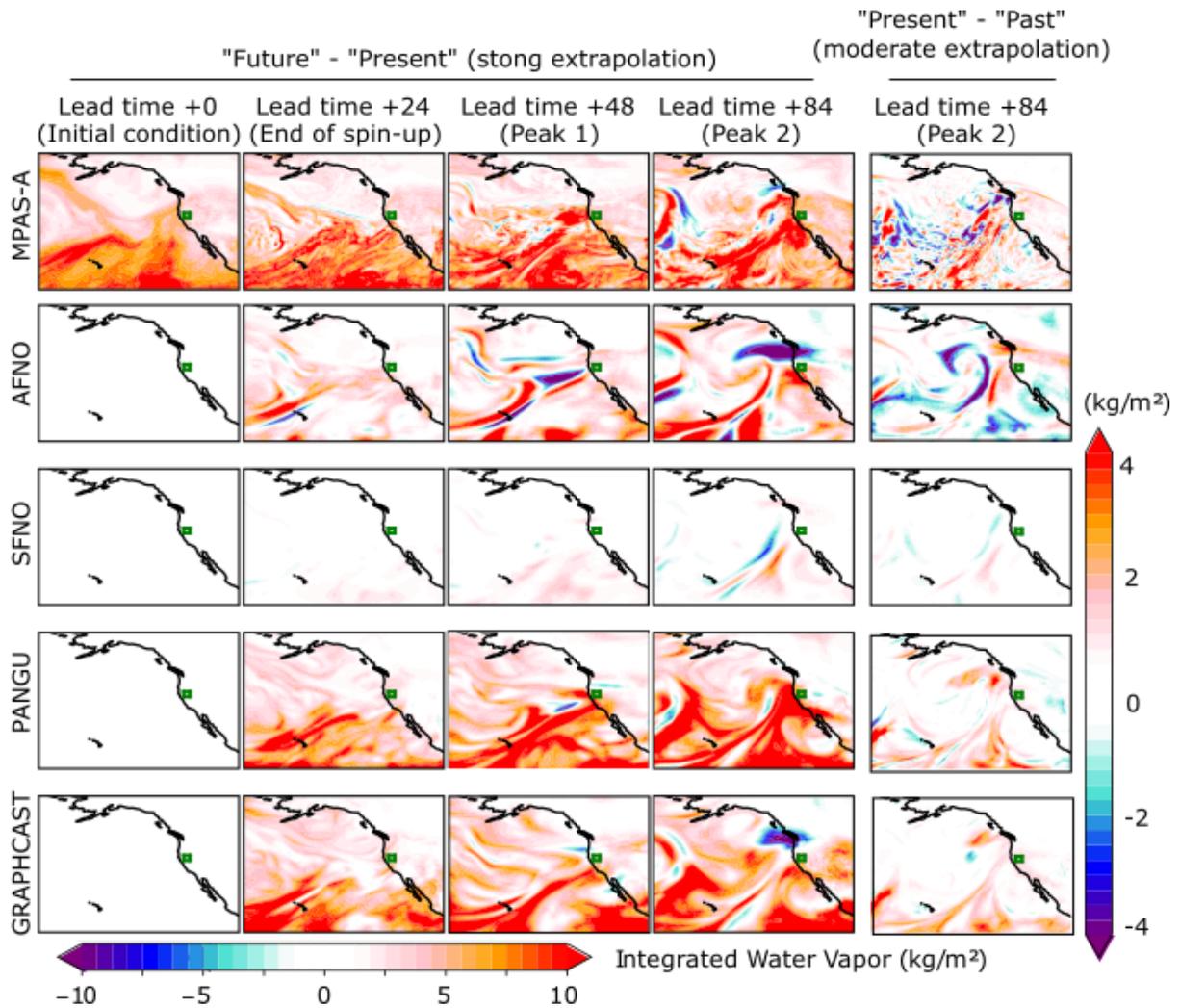

Fig. 4. "Future" minus "present" simulations (strong extrapolation) of MPAS-A and the AI data-driven model of Integrated Water Vapor (IWV) over the North Central and North Eastern Pacific Ocean and Western coast of the United States (US) for four different dates: February 5th at 12 UTC, February 6th at 12 UTC (lead time of 24 hours), February 7th at 12 UTC (lead time of 48 hours), and February 9th at 0 UTC (lead time of 84 hours), representing the initial condition, the end of MPAS-A spin up in Michaelis et al., 2022, the first peak, and the second peak of the AR, respectively. "Present" minus "past" simulations (moderate extrapolation) are displayed in the last column for the second peak. The colorbar on the bottom- and right-side describe the magnitudes for the strong, and moderate extrapolation cases, respectively.

*d. On the initialization module*

Model simulations run with the initialization module (see Section 4) for the different AI-models are shown in Figure 5 at a lead time of 84 hours for the "future" scenario. "Future" minus "present" simulations are displayed for the air temperature at 850 hPa (first two rows) and the IWV (last two rows). Values for simulations without the initialization module are also included in the Figure for comparison purposes, together with the MPAS-A ones, which are used as reference. Therefore, the goal is to measure the impact of the

File generated with AMS Word template 2.0

initialization module on the extrapolation response of the AI-models, given the initial condition now presents geopotential and pressure fields which are balanced with the new thermodynamic situation. Both AFNO and SFNO show important changes when run with the initialization module. They exhibit higher values for temperature and IWV than those simulated without the module. Particularly, the SFNO fields are well-aligned with MPAS-A capturing impressively the spatial pattern. In spite of this, SFNO still shows a certain lack of extrapolation, especially over the ocean. The influence of the geopotential at initial time –which are the main fields that were modified in the initialization module– on the SFNO and AFNO simulations is consistent with recent findings in the literature (Baño-Medina et al., 2024c). Differently, the results for Pangu and Graphcast are not so encouraging, since they exhibit larger differences as compared to MPAS-A than the simulations run without the initialization module, worsening the extrapolation skill.

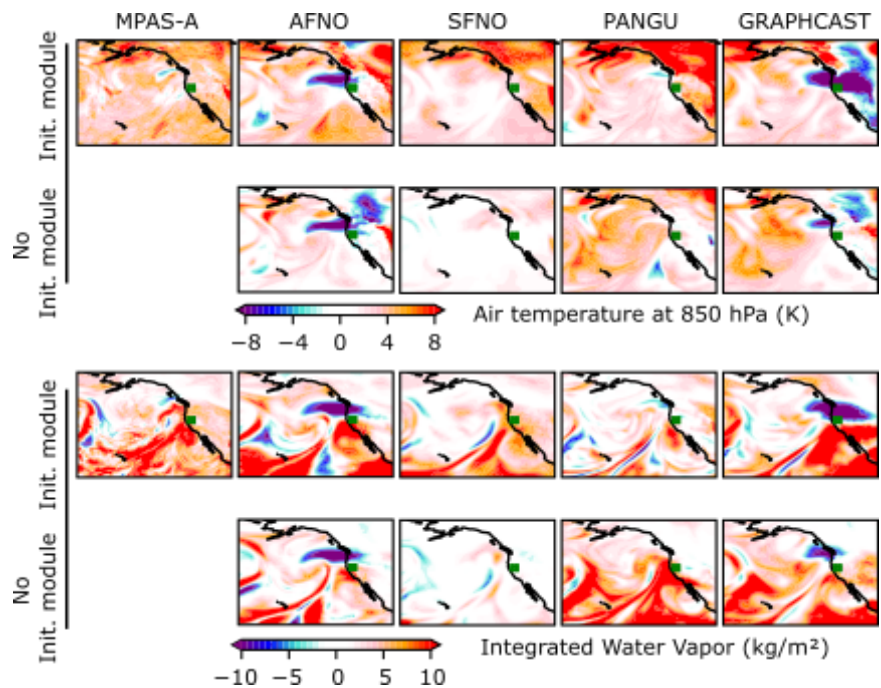

Fig. 5. "Future" minus "present" simulations for the MPAS-A, AFNO, SFNO, Pangu, and Graphcast for the air temperature at 850 hPa (first two rows), and the integrated water vapor (last two rows) on February 9th at 0 UTC (lead time of 84 hours). Simulations are run with (rows 1, and 3), and without (rows 2, and 4) the initialization module. Note that for the MPAS-A the latter case does not apply.

*e. On the attribution skill at the Oroville dam*

IWV differences over the Oroville dam domain (yellow box in Figure 2) between the "future" and the "present" (strong extrapolation case, first row), and the "present" and the

File generated with AMS Word template 2.0

"past" (moderate extrapolation case, second row) simulations are displayed in Figure 6. Models appear with different colors, and again the MPAS-A simulations are used as an extrapolation reference for the AI-models. The MPAS-A projects higher values of IWV for the "future" simulation throughout the entire AR episode, reaching 4 and 6 kg/m² increases for the first and second peaks, respectively; MPAS-A attributes about 1 kg/m² (5–6%) of the "present" values to the anthropogenic influence (i.e., comparing "present" to "past"). These increased values of moisture in the AR also result in more precipitation as shown in Michaelis et al. (2022), which could make the difference between a beneficial AR (i.e., ARs producing volumes of water manageable by the water management services) and a hazardous one (i.e., ARs causing devastating impacts in the area). In the case of the AI models, for the moderate extrapolation case, AFNO, Pangu and Graphcast show magnitudes of IWV differences which are comparable to MPAS-A, while SFNO produces minimal increases with respect to the "past" simulation. For the strong extrapolation case, the AI data-driven models produce larger increments of IWV with the exception of SFNO, but they are smaller compared to MPAS-A. This finding is in line with the results from Figures 2, and 3, where extrapolation deficiencies over the Pacific and Western US were also observed. Nevertheless, the AI data-driven models are able to reproduce the MPAS-A signatures of increased IWV throughout the period and, in some cases, with comparable magnitudes. Results for the remainder of the simulations can be found in Appendix C.

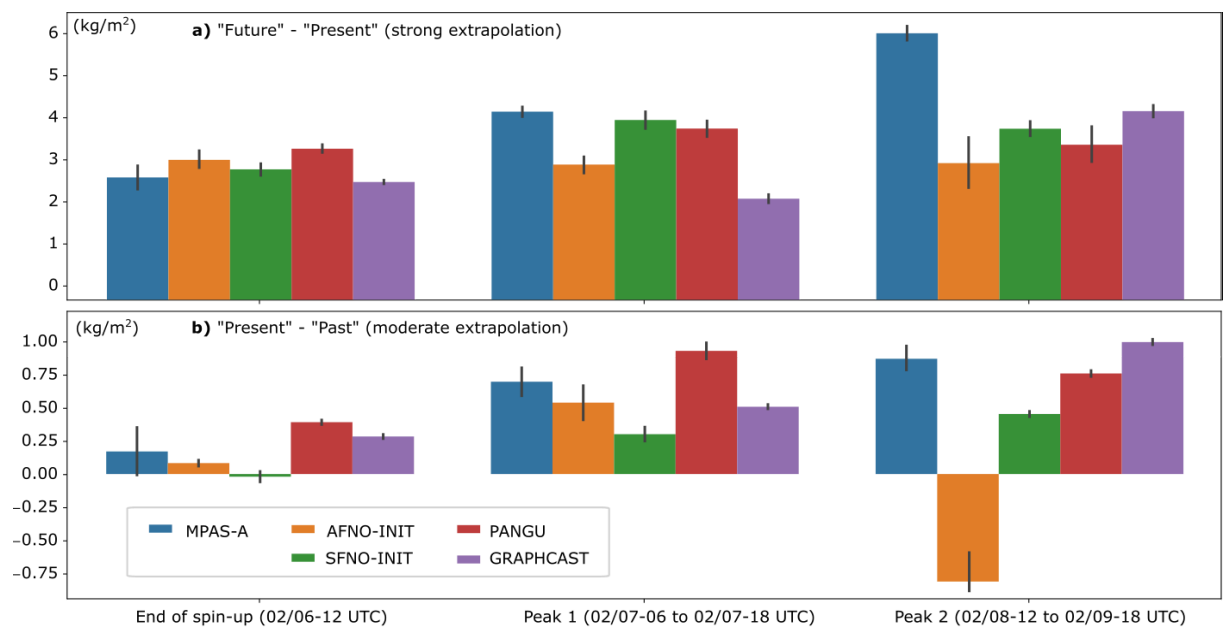

File generated with AMS Word template 2.0

Fig. 6. a) "Future" minus "present" simulations (strong extrapolation) of MPAS-A and the AI data-driven model of Integrated Water Vapor (IWV) over the Oroville's dam (see yellow box in Figure 1) for three different dates: February 6th at 12 UTC (lead time of 24 hours), February 7th at 12 UTC (lead time of 48 hours), and February 9th at 0 UTC (lead time of 84 hours), representing the end of MPAS-A spin up in Michaelis et al., 2022, the first peak, and the second peak of the AR, respectively. b) "Present" minus "past" simulations (moderate extrapolation). Models are displayed in different colors. Bars represent the spread, i.e., min-max interval of values covered by the different grid points in the Oroville dam domain.

*f. On the potential of large ensembles for extreme events and climate attribution*

Figure 7 shows the time series of the Oroville AR as described by the EnAFNO (540-members) and MPAS-A (21-members) probabilistic systems for the (from top to bottom) "present", "future", and "past" simulations. Following ERA5, the characteristic two peaks of the Oroville AR are clearly evident, with the second one showing higher values and a longer duration. There is also a minimum between the peaks, coinciding with the time when the two ARs merge into a single, and stronger, AR over the Pacific Ocean (Figure 1). Both MPAS-A and EnAFNO are able to reproduce these main characteristics of the event under "present" conditions, with some differences. For example, the ensemble mean of EnAFNO fails to capture the intensity of the second peak while MPAS-A overestimates IWV during the gap in AR conditions between the two peaks. The difference of ensemble sizes and thus, the IWV range covered at each time-step, is notable between MPAS-A, and EnAFNO. EnAFNO spread clearly increases as a function of lead time, in line with the non-linear and chaotic nature of the atmosphere. The large number of members within EnAFNO allows the ensemble system to assign probabilities to values that are in alignment with the ERA5 ground-truth over the valley and the second peak, whereas the MPAS-A ensemble spread is not enough to capture the ERA5 values between IWV peaks. For the "future" simulations both MPAS-A and EnAFNO project increased values of IWV over the peaks, in line with Figure 6. The time series nicely illustrates the spin-up of IWV as a function of lead time, shifting from the "present" IWV values at initial time to higher values surpassing the ERA5 reference and its own forecast in the "present" scenario. This exemplifies the temperature-moisture link shown in Figure 4, and the physical consistent response of this network. However, as mentioned before, the IWV increments are not as large as those produced by MPAS-A, which is consistently greater than the ERA5 reference throughout the entire AR episode. For the "past" simulations both models show slightly similar results to their "present" ones. Notably, the large number of members within EnAFNO provides a considerably wider range of IWV values for both the "future" and "past" simulations, and thus larger uncertainties, or possibilities, for the attribution of the event.

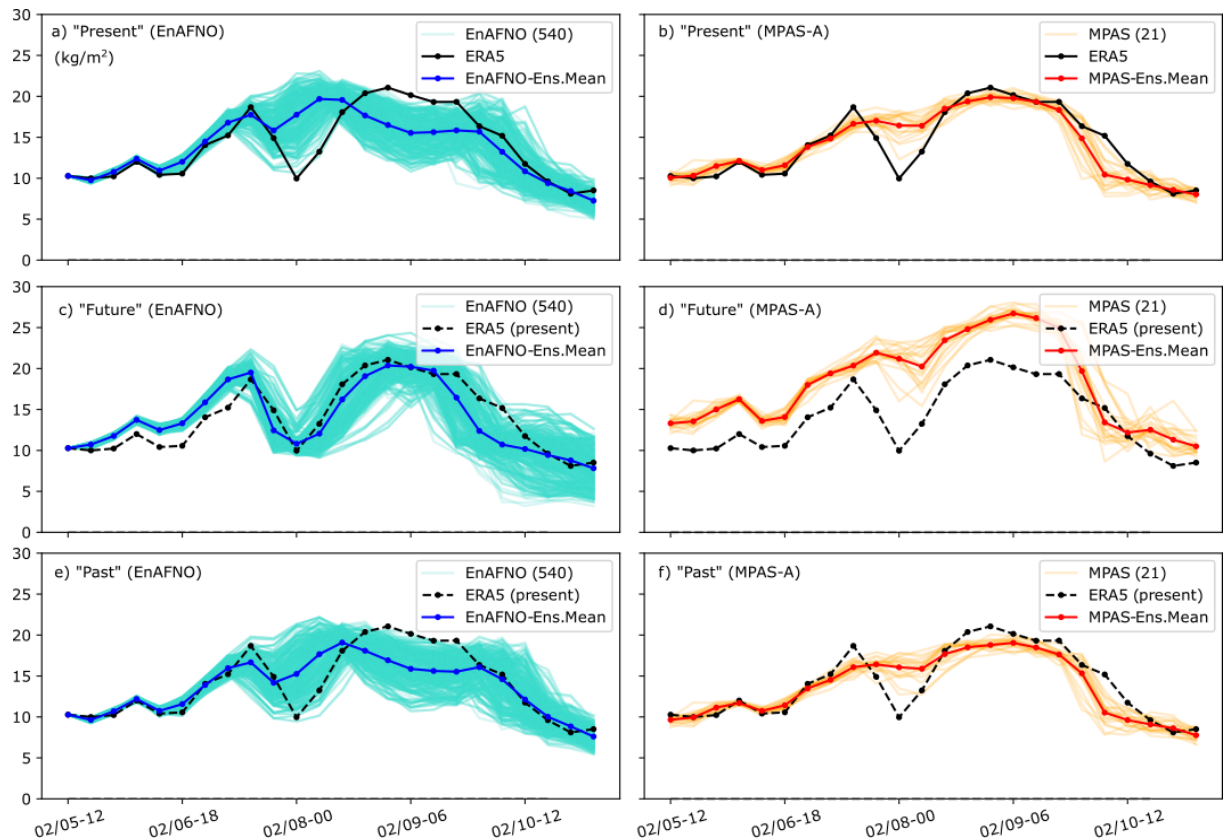

Fig. 7. Time series of IWV during the Oroville AR as described by the EnAFNO (540-members) and MPAS-A (21-members) probabilistic systems, for the (from top to bottom) "present", "future", and "past" simulations. ERA5 is represented in black.

To illustrate the potential of large ensembles for climate attribution, we performed a t-test for the first peak of the AR (April 7th, 12 UTC) with a significance level of 0.05 where the null hypothesis is that the "present" minus "past" simulations are equal to 0, i.e., there is no human fingerprint on this event. The mean values of these differences for EnAFNO and MPAS-A are 1.08 $kg/m^2$, and 0.56 $kg/m^2$, respectively; whereas the standard deviations are 0.4 $kg/m^2$, and 0.36 $kg/m^2$, respectively. P-values for EnAFNO and MPAS-A are 0.006 and 0.16, which means that EnAFNO present-day to pre-industrial attribution response at this date is statistically significant where MPAS-A one is not.

## 6. Discussion

The analysis presented provides an examination of the key properties required to conduct attribution studies with AI-models: ability to forecast an extreme event, the extrapolation ability, the relationship between temperature and moisture (related to energy conservation), the initialization module, and the potential of large ensembles. The conclusions derived from this study are exclusively for the Oroville AR; future work, including other types of extreme events such as heat waves or cyclones, will allow broader conclusions about the suitability of the current generation of AI data-driven models for climate attribution. A few examples can be found in a concurrent analysis to this work with the SFNO model for the moderate extrapolation case (Jiménez-Esteve et al., 2024).

Two extrapolation cases are presented—strong and moderate—building on the temperature "delta" fields from a CMIP5 GCM ensemble under RCP8.5 and pre-industrial scenarios relative to a historical period, respectively. The resulting simulations, three for each model, are compared to measure the human fingerprint on the magnitude of the Oroville AR (moderate extrapolation case) and to elucidate its potential magnitude in the "future" (strong extrapolation case). To evaluate the extrapolation ability of AI data-driven models, MPAS-A simulations are used as a "pseudo-reality", following well-established practices in the climate change literature (e.g., Vrac et al., 2017). Note that conclusions drawn from this evaluation procedure entails risks, since dynamical models also present uncertainties in their climate change simulations (e.g., Boé et al., 2020), although MPAS-A was specifically designed to model such events in detail. The AI-models are able to preserve a large fraction of the temperature signal throughout the entire AR episode. Pangu and Graphcast are able to do this without adjusting the geopotential –based on the hydrostatic balance in the initial condition field– while AFNO and SFNO extrapolation skill clearly depends on this component of the initialization module due to the strong influence of the geopotential in these models (Baño-Medina et al., 2024c), which prevents them to extrapolate to the new thermodynamic atmospheric state. Nevertheless, the evolved temperature signal for all AI models during the second peak present smaller changes than MPAS-A, and thus suggests poor extrapolation skill, especially for the strong extrapolation case. Pre-training the models with climate simulations has been a successful practice to improve the extrapolation ability of certain statistical downscaling (Doury et al., 2023; Boé et al., 2023; Baño-Medina et al., 2024b), and seasonal forecasting (Gibson et al., 2021) models, by providing them with samples from different scenarios during training. A similar practice could help this generation of AI data-driven models to better extrapolate to climate change regimes.

Modeling the interactions between atmospheric variables is not explicitly tasked in the neural network. Therefore, the AI models might not necessarily produce a physically consistent response to perturbations in an atmospheric variable. However, some cases of study have shown how these models might actually be producing forecasts based on physical mechanisms (Hakim & , Baño-Medina et al., 2024c). This is key in climate attribution, since links between the temperature fields and other variables (e.g., circulation, winds, moisture) must be properly captured. Here, the moisture-temperature link was examined due to its relevance to the Oroville AR and resultant precipitation. Graphcast, Pangu, and, to a lesser degree, AFNO, produced increased levels of moisture in line with the temperature increments given by the "delta" fields, demonstrating the ability to rapidly shift the IWV values towards magnitudes consistent with the temperature fields, similar to the spin-up of dynamical models. Again, SFNO required the initialization module to adjust the geopotential, so it is in hydrostatic balance, to produce consistent increments of IWV with the temperature "deltas". Moreover, important similarities in the spatial patterns between the MPAS-A and the AI data-driven models response were also identified, highlighting that the moisture-temperature link learned is to some degree built on physics, and thus demonstrating their ability to produce physically consistent responses to "climate-change" perturbations. However, the strong dependence of SFNO on the geopotential suggests potential covariability with the temperature variables, which can hinder reliability. Also, adjusting the hydrostatic balance and consequently modifying the geopotential fields in Pangu and Graphcast seem to worsen their extrapolation skill, which could be also an indicator of potential improvements on the geopotential-temperature-moisture inter-variable links. To properly capture these atmospheric links in the neural network coefficients is key to gain confidence on these tools for attribution analysis. Loss functions including expressions describing the interactions between atmospheric variables could help to better model this aspect. A few physics-informed neural networks have been developed for atmospheric studies (e.g., González-Abad et al., 2023), but are yet to be explored for AI data-driven weather models. Other alternatives are to train additional networks emulating the initialization modules of dynamical models, such as the NN-INIT model developed here. Moreover, this type of networks can even avoid assumptions in the modeling of the atmosphere, such as the hydrostatic balance assumption, that might not be entirely accurate and could lead to errors in the simulations. However, NN-INIT comes with its own sources of error, since this network is tasked to complete the remaining atmospheric fields of the initial condition given a limited number of variables at the same time.

The aforementioned properties led to similar attribution values of IWV for the Oroville AR compared to MPAS-A, and projected an even more extreme event in the future climate than in the present one, also in line with the dynamical model. These results exemplify the potential and the current limitations of this generation of AI models to conduct attribution analysis of extreme events. The short times required to produce the simulations can help increase the number of studies analyzed, and produce better representations of the uncertainty, in near-real time when public attention is heightened. Moreover, the results herein could also be relevant for any climate change-based study, and can guide development of the next generation of AI data-driven models beyond the weather time scales.

*Acknowledgments.*

This work was supported by the California Department of Water Resources Atmospheric River Program Phase III and Phase IV (Grants 4600014294 and 4600014942, respectively) and U.S. Army Corps of Engineers (USACE) Forecast Informed Reservoir Operations Phase 2 Award (USACE W912HZ192). A.S. was partly supported by the National Aeronautics and Space Administration (Grant 80NSSC22K0926). The CMIP5 GCM ensemble mean data for the future scenario and interpolation codes used in this study were provided by Chunyong Jung at Argonne National Laboratory.

*Data Availability Statement.*

Graphcast, Pangu, SFNO, and AFNO models are available through the ecmwf-ai Github (https://github.com/ecmwf-lab/ai-models) supported by ECMWF. The code for the EnAFNO ensemble system is available at the Center for Western Weather and Water Extreme Github repository (CW3E, https://github.com/CW3E/EnAFNO). Code to reproduce the results of the manuscript can be found also in CW3E's Github repository: https://github.com/CW3E/AI-attribution. The data and the NN-INIT model are publicly available at the University of California San Diego (UCSD) repository (Baño-Medina et al., 2024d).

# Appendix A

## Hydrostatic balance component of the initialization module

The following integration scheme (from bottom to top) is used to adjust the geopotential fields ($z$) to be in hydrostatic balance with the modified virtual temperature fields ($Tv$) at initial time, which are functions of the temperature and relative humidity fields at each pressure level ($l$). $R$ is the ideal gas constant, for air $R=287\ J/kg.K$.

$$z_{l+1} = z_l - ln(p_{l+1} - p_l) \times R \times (Tv_{l+1} + Tv_l) \times 0.5$$

# Appendix B

## Integration of IWV for Pangu and Graphcast

The following integration scheme (from bottom to top) is used to compute the Integrated Water Vapor (*IWV*) for Pangu and Graphcast, given this variable is not included in their model formulation. Both models contain values for the specific humidity ($q$) at 13 pressure levels ($p$): 1000, 925, 850, 700, 600, 500, 400, 300, 250, 200, 150, 100, and 50 hPa.

$$IWV = 0.5 \times (q_l + q_{l+1}) \times (p_l - p_{l+1})$$

# Appendix C

## On the attribution skill at the Oroville dam

Figure C shows IWV differences over the Oroville dam domain (yellow box in Figure 2) between the "future" and the "present" (strong extrapolation case, first row and third row), and the "present" and the "past" (moderate extrapolation case, second and fourth row) simulations are displayed in Figure C1 for simulations including (rows 1-2) and not including (rows 3-4) the initialization module. Models appear with different colors, and again the MPAS-A simulations are used as an extrapolation reference for the AI-models.

File generated with AMS Word template 2.0

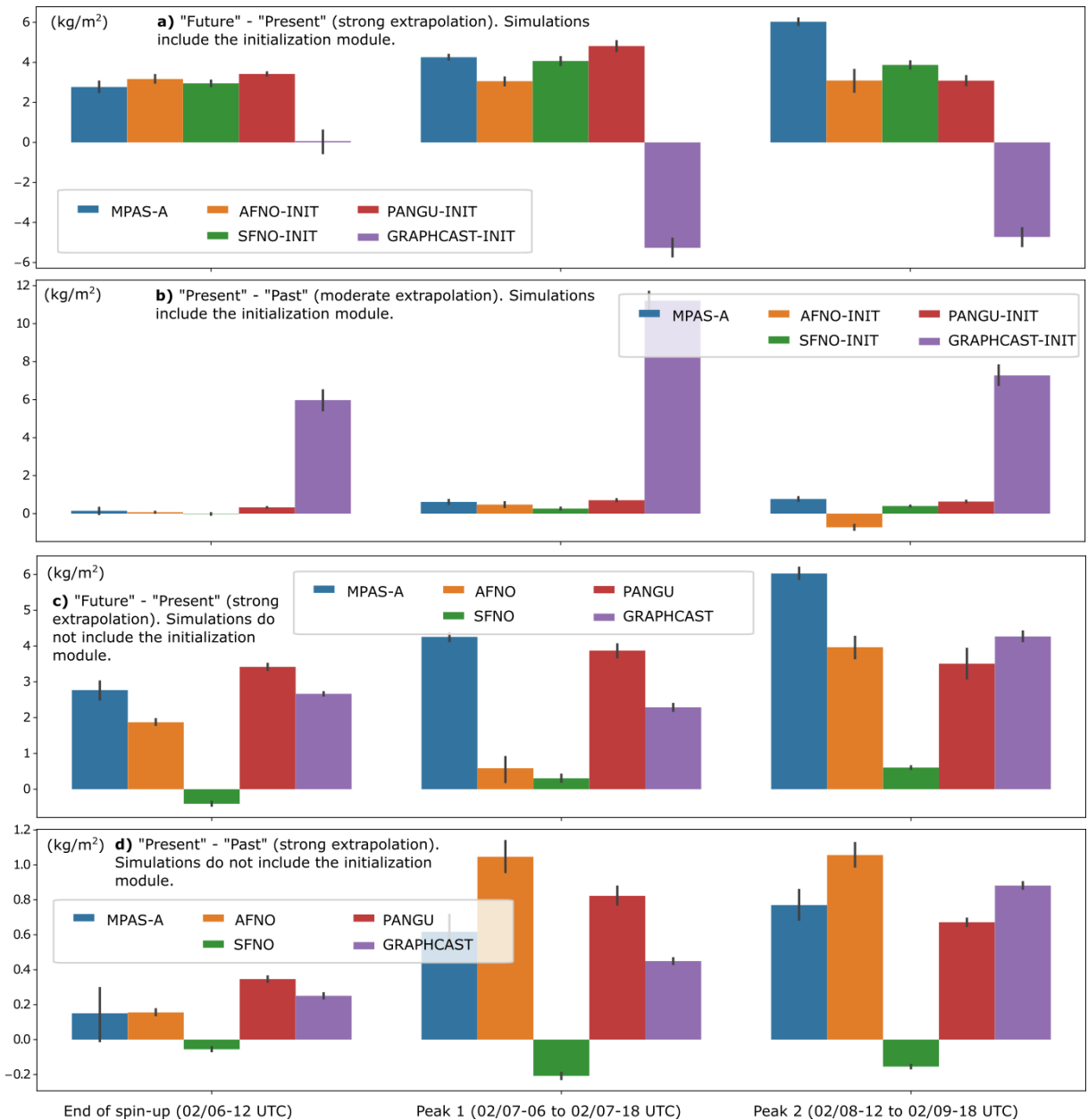

Fig. C. a) "Future" minus "present" simulations (strong extrapolation) of MPAS-A and the AI data-driven model of Integrated Water Vapor (IWV) over the Oroville's dam (see yellow box in Figure 1) for three different dates: February 6th at 12 UTC (lead time of 24 hours), February 7th at 12 UTC (lead time of 48 hours), and February 9th at 0 UTC (lead time of 84 hours), representing the end of MPAS-A spin up in Michaelis et al., 2022, the first peak, and the second peak of the AR, respectively. b) "Present" minus "past" simulations (moderate extrapolation). Models are displayed in different colors. Bars represent the spread, i.e., min-max interval of values covered by the different grid points in the Oroville dam domain. Panels c) and d) are analogous to a) and b), respectively, but simulations do not include the initialization module, whereas this module is included in the simulations of Panels a) and b).

Frei, C., Schöll, R., Fukutome, S., Schmidli, J., & Vidale, P. L. (2006). Future change of precipitation extremes in europe: Intercomparison of scenarios from regional climate models. Journal of Geophysical Research: Atmospheres, 111(D6).

Gao, Y., Lu, J., Leung, L. R., Yang, Q., Hagos, S., & Qian, Y. (2015). Dynamical and thermodynamical modulations on future changes of landfalling atmospheric rivers over western north america. Geophysical Research Letters, 42(17), 7179–7186.

Gershunov, A., Shulgina, T., Clemesha, R. E., Guirguis, K., Pierce, D. W., Dettinger, M. D., ... & Ralph, F. M. (2019). Precipitation regime change in Western North America: the role of atmospheric rivers. *Scientific reports*, *9*(1), 9944.

Gibson, P. B., Chapman, W. E., Altinok, A., Delle Monache, L., DeFlorio, M. J., & Waliser, D. E. (2021). Training machine learning models on climate model output yields skillful interpretable seasonal precipitation forecasts. *Communications Earth & Environment*, *2*(1), 159.

González-Abad, J., Hernández-García, Á., Harder, P., Rolnick, D., & Gutiérrez, J. M. (2023). Multi-variable hard physical constraints for climate model downscaling. In *Proceedings of the AAAI Symposium Series* (Vol. 2, No. 1, pp. 62-67).

Goodfellow, I., Bengio, Y., & Courville, A. (2016). Deep learning. MIT press.

Guibas, J., Mardani, M., Li, Z., Tao, A., Anandkumar, A., & Catanzaro, B. (2021). Adaptive fourier neural operators: Efficient token mixers for transformers. arXiv preprint arXiv:2111.13587.

Guo, X., Huang, J., Luo, Y., Zhao, Z., & Xu, Y. (2017). Projection of heat waves over china for eight different global warming targets using 12 cmip5 models. Theoretical and applied climatology, 128, 507–522.

Hakim, G. J., & Masanam, S. (2023). Dynamical tests of a deep-learning weather prediction model. arXiv preprint arXiv:2309.10867.

Hazeleger, W., van den Hurk, B. J., Min, E., van Oldenborgh, G. J., Petersen, A. C., Stainforth, D. A., . . . Smith, L. A. (2015). Tales of future weather. Nature Climate Change, 5(2), 107–113.28
File generated with AMS Word template 2.0

File generated with AMS Word template 2.0